\documentstyle[psfig,12pt]{article}

\def\beq{\begin{equation}}
\def\eeq{\end{equation}}

\begin{document}

\begin{titlepage}

\begin{centering}

\vspace*{3cm}

{\Large\bf Discrete light-cone quantization on a twisted torus}

\vspace*{1.5cm}

{\bf S.S.~Pinsky}
\vspace*{1.0cm}

{\sl Department of Physics\\
Ohio State University\\ Columbus, OH 43210, USA}\\

\vspace*{1.5cm}

{\large Abstract}

\end{centering}

\vspace*{.8cm}

Recently it has been demonstrated by Dienes and Mafi that the physics of
toroidal compactified models of extra dimensions can depend on the shape 
angle of the torus.  Toroidal compactification has also recently been used 
as a regulator for numerical solutions of supersymmetric field theories in 
2+1 dimensions. The question is: does the shape angle of the torus also 
affect the physics in this situation?  Clearly a numerical solution should 
be independent of the shape of the space on which we compactify, at least
when the regulator is removed. We show that, for standard discrete light-cone
quantization with transverse parity invariance, toroidal compactification is
only allowed for a specific set of shape angles and for that set of shape 
angles the numerical solutions are unchanged.

\vspace{0.5cm}


\vfill

\end{titlepage}
\newpage
\section{Introduction}
Using a method we call Supersymmetric Discrete Light-Cone Quantization (SDLCQ)
we have been able to solve a number of supersymmetric theories in 2+1
dimensions~\cite{hpt,Haney:2000tk,Antonuccio:1999zu}.  This method, which
is an extension of DLCQ~\cite{pb85}, exactly
preserves supersymmetry and requires no renormalization in
$2+1$ dimensions. 

DLCQ is a  numerical method for solving quantum field theory that is actually
the combination of two very well known ideas.  The first idea, light-cone (LC)
quantization, originally proposed by Dirac in 1949~\cite{dir49},
points out that one can evolve a system with operators other than $P^0$. When
the system is evolved with the operator $P^-=(P^0-P^1)/\sqrt{2}$ this leads, 
when quantized, to LC quantization. In LC quantization one replaces
$(x^0,x^1,x^\perp)$ by $(x^+,x^-,x^\perp)$  where 
$x^\pm = (x^0 \pm x^1)/\sqrt{2}$.  The metric is implicitly defined by
$x^\pm=x_\mp$ and $x^\perp=-x_\perp$. In general $x_\perp$ can have any
number of components, but here we will be considering only one transverse
coordinate $x_2$. In this system $x_-$ is the LC time, and $P^-$ is the LC
Hamiltonian. In DLCQ one regulates the system by
putting it in a LC spacial box with boundary conditions on $x_\perp$ and 
$x_+$, which gives rise to discrete momentum modes in $P^+$ and $P^\perp$. 
The modes are formulated in Fock space.  Truncation then turns
the quantum field theory into a finite-dimensional numerical problem. 
A detailed review of DLCQ can be found in~\cite{bpp98}.

\begin{figure}
\hspace{1.5in}
\psfig{figure=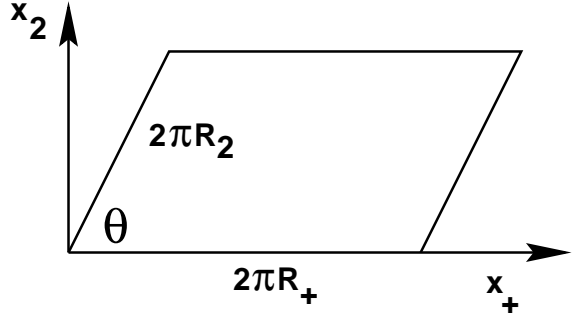,width=7.5cm,angle=0} 
\caption{General two-dimensional torus with shape angle $\theta$.}
\label{torus}
\end{figure}

In the context of extra-dimensional physics, Dienes and
Mafi~\cite{dienes,dienes2} considered compactification on a
generalized torus, shown in Fig.~\ref{torus}, which contains a
shape angle $\theta$. Dienes and Mafi showed that the properties of the
Kaluza--Klein particle in an extra-dimensional field theory depend on this
shape angle~\cite{dienes,dienes2}.

This naturally leads to the question: when we formulate a (2+1)-dimensional
supersymmetric field theory on a torus and solve it using DLCQ, will the 
physics depend on the shape angle $\theta$ as well? The difference between
a truly extra-dimensional theory and DLCQ is that
the cylinder in DLCQ is introduced as a regulator for the field
theory rather than as a fundamental part of the theory. One could think of 
the compactification in $x_2$ as a model for a true extra dimension, but we
will not consider that here. If we were to find that the results depend on
the shape angle, it would surely put in question this method of regulating
(2+1)-dimensional DLCQ theories.

The torus with shape angle $\theta$ is shown in Fig.~\ref{torus}.  The
periodicities of the torus take the form
\begin{eqnarray}
     && \cases{  x_+ ~\to~ x_+ + 2\pi R_+\cr
                     x_2 ~\to~ x_2~,\cr}  \nonumber\\
         && \cases{  x_+ ~\to~ x_+ +2\pi R_2 \cos\theta\cr
                     x_2 ~\to~ x_2+2\pi R_2 \sin\theta~.\cr}
\label{torusdef}
\end{eqnarray}
The conventional or rectangular torus corresponds to $\theta =\pi/2$. 
In discussing the generalized torus it is conventional to introduce
the complex quantity $\tau$,
\begin{equation}
\tau\equiv  \frac{R_2}{R_+}e^{i\theta}
    =\tau_1+i\tau_2=\frac{R_2}{R_+}\cos\theta +i\frac{R_2}{R_+}\sin\theta\,.
\end{equation}
It is also conventional within this context to normalize the scale by
taking the side of the torus along the horizontal axis to be of length one,
i.e.\ $2\pi R_+=1$. In this form the torus is represented by $\tau$ in the
complex $\tau$ plane. It can be shown that the torus has an invariance, 
generally called a modular invariance~\cite{conform}.  One of these
modular transformations, which will play a key role here, is
$\tau_1'=\tau_1 +1$. 

The periodic functions that replace the simple exponential are
\begin{equation}
     \exp\left( i {n_+\over R_+} \left[x_+ - {x_2\over
\tan\theta}\right] 
                 ~+~ i{n_2\over R_2} {x_2\over \sin\theta} \right)~.
\label{KKfuncts}
\end{equation}
For the case where $n_+$ and $n_2$ are (odd half) integers, the expression
in Eq.~(\ref{KKfuncts})
is a (anti-)periodic function.  We will focus on periodic boundary conditions 
because those are required in SDLCQ, but the conclusions are also valid for 
anti-periodic boundary conditions.

In Section 2 we will carefully define the standard DLCQ formulation of a free
scalar field theory in 2+1 dimensions. We will then ask if this theory with 
the same cutoffs can be defined on the torus with a shape angle. We will find 
that the theory can only be defined on a subset of tori. That is, only some 
shape angles are allowed. We will then show that for this subset of allowed 
shape angles the physics is unchanged. In Section 3 we will show that this
result carries over to the SDLCQ formulation of supersymmetric Yang--Mills
(SYM) theory in 2+1 dimensions.

\section{DLCQ} \label{sec:dlcq}
We start by considering the DLCQ formulation of the theory for
a free massive boson in $2+1$ dimensions.  The Lagrangian is
\begin{equation}
{\cal L}= \frac{1}{2}\partial_\mu \phi\partial^\mu \phi-\frac{1}{2} m^2
\phi^2\,.
\end{equation}
The LC Hamiltonian for this theory is
\begin{equation}
P^-=\int_{-\infty} ^\infty dx_+dx_2 (\frac{1}{2}\partial_2 \phi \partial^2
\phi  +\frac{1}{2}m^2 \phi^2)\,.
\end{equation}
By the phrase ``standard DLCQ'' we mean precisely this set of interactions,
with all of its symmetries, and a discrete set of longitudinal momentum modes, 
with a longitudinal resolution $K$ and cutoffs that respect the symmetries.

We now want to quantize this theory on a torus with a shape angle but 
otherwise follow this standard DLCQ procedure. It is straightforward to 
quantize this theory using the periodic functions in Eq.~\ref{KKfuncts}. 
We define creation and annihilation operators that satisfy the standard
commutation relations
\begin{equation}
 \left [ A(n_+,n_2), A^\dagger(m_+,m_{2}) \right]=\delta_{n_+,m_+}
\delta_{n_2,m_2}.
\end{equation}
In terms of these operators the second-quantized field takes the form
\begin{equation}
\phi(x) = \frac{1}{2\pi\sqrt{R_2 \sin\theta}}\sum_{n_+=0}^\infty
\sum_{n_2=-\infty}^\infty [e^{-i(\frac{n_+}{R_+}Z_+ +
\frac{n_2}{R_2}Z_2)} A(n_+,n_2)
+ e^{i(\frac{n_+}{R_+}Z_+ + \frac{n_2}{R_2}Z_2)}
A^\dagger(n_+,n_2){]}\,,
\end{equation}
where
\begin{eqnarray}
Z_2 &=& \frac{x_2}{\sin\theta}\,, \nonumber \\
Z_+ &=& x_+ - \frac{x_2}{\tan\theta}\,.
\end{eqnarray}
We define the Hamiltonian in momentum
space by integrating over the torus. To actually do the integrals it is
convenient to change to the variables $Z_+$ and $Z_2$, because
\begin{equation}
\int_{\rm torus} dx_+ dx_2 =
     \sin\theta\int_0^{2\pi R_+}dZ_+\int_0^{2\pi R_2}dZ_2\,.
\end{equation}
The Hamiltonian then takes the form
\begin{equation}
P^-=\sum_{n_+=1}^\infty\sum_{n_2=-\infty}^\infty \frac{R_+}{2n_+}
[ m^2 +\frac{n_2^2}{R_2^2 \sin\theta^2} (1-\frac{n_+}{n_2}\tau_1
)^2] A^\dagger(n_+,n_2)A(n_+,n_2)\,.
\end{equation}

Following the standard DLCQ procedure, we will now define the Fock basis on 
a torus with a shape angle. In standard DLCQ we
use transverse boost invariance to work in the frame where the total 
transverse momentum is zero. We will refer to this frame as the total 
transverse momentum  center of momentum (TMCM) frame.  
In standard DLCQ the total longitudinal momentum $P^+$ is given by
$P^+=\frac{K}{R_+}$, the $i$-th particle has a longitudinal momentum
$n_{(i)+}/R_+$, and the sum of the integers $n_{(i)+}$ is just $K$.
We follow the same procedure on the torus with a shape angle. The 
single-particle Fock state is
\begin{equation}
|\psi_1 \rangle = A^\dagger(n_+,n_2) |0\rangle\,.
\end{equation}
For this state we have $n_+=K$ and a total transverse momentum
\begin{equation}
P_2=\frac{1}{R_2\sin\theta}(n_2-n_+\frac{R_2}{R_+}\cos\theta)\,.
\end{equation}
In the TMCM frame this transverse momentum is zero, and therefore
\begin{equation}
n_2=K\tau_1\,.
\end{equation}
It will be convenient to define the integer $n\equiv K\tau_1$.  Then the above 
equation can be simply written $n_2=n$, and we conclude that
$\tau_1=\frac{n}{K}$ must be a rational number. This is the first restriction 
we find on the allowed tori. Applying the Hamiltonian to this state we find
\begin{equation}
2\frac{K}{R_+}P^-|\psi_1\rangle = m^2 |\psi_1 \rangle\,,
\end{equation}
as expected. The physics of this one-particle state is unchanged on a torus 
with a shape angle, provided that $\tau_1$ is a rational number.

Now let us consider a two-particle Fock state. A general state with longitudinal resolution
K has the form
\begin{equation}
|\psi_2\rangle=\sum_{n_+, n_2,n_2'}
f_2(n_+,n_2,n_2')A^\dagger(n_+,n_2)A^\dagger(K-n_+,n_2')|0\rangle\,.
\end{equation}
As we will see, the sums on $n_2$ and $n_2'$ are not independent. The 
transverse momentum of the two particles in this state are
\begin{eqnarray}
k_2&=&\left(\frac{n_2}{R_2 \sin\theta}-\frac{n_+}{R_+\tan\theta}\right)\,,\\
k_2'&=&\left(\frac{n_2'}{R_2 \sin\theta}-\frac{K-n_+}{R_+\tan\theta}\right)\,.
\label{k2}
\end{eqnarray}
In the TMCM frame the sum $k_2+k_2'$ is zero, and we find
\begin{equation}
n_2+n_2'=K\frac{R_2}{R_+}\cos\theta=K\tau_1=n\,.
\end{equation}
Again the restriction that $\tau_1$ must be a rational number appears.
It is straightforward to generalize this to higher Fock states, and we find 
that $\sum_i n_{(i)2}=n$. The general form of the two-particle state is now
\begin{equation}
|\psi_2\rangle=\sum_{n_+=1}^{K-1} \sum_{n_2}
f_2(n_+,n_2)A^\dagger(n_+,n_2)A^\dagger(K-n_+,n-n_2)|0\rangle\,.
\end{equation}
This theory is invariant with respect to transverse parity, 
$k_2 \rightarrow -k_2$. It is easy to see that in terms
of $n_2$ this transformation is $n_2 \rightarrow -n_2 +n_+2\tau_1$. It can 
also be shown that this transformation works for all higher states as well. 
For this to be a symmetry of the discrete theory, we conclude that
$2n_+\tau_1$ must be an integer.  

Also in DLCQ we must truncate the sum on
$n_2$ to make the problem numerically solvable. We impose the conventional DLCQ
cutoff, a symmetric cutoff in $k_2$, to preserve transverse
parity~\cite{hpt,Haney:2000tk,Antonuccio:1999zu}, which can be imposed 
independently of the shape angle.
We will take the lower and upper cutoffs of the $n_2$ sum to be $-T_l$ and
$T_u$. These translate into upper and lower cutoffs on $k_2$, and for 
these to be symmetric we must have
\begin{equation}
T_u=T_l+2 n_+ \tau_1\,.
\end{equation}
Again we find that $2n_+\tau_1$ must be an integer.

This condition on $2n_+\tau_1$ leaves us with two allowed cases to consider, 
that $\tau_1$ is an integer or a half integer. Using the  modular transformation
$\tau_1'=\tau_1 +1$,  we see that all tori are equivalent to tori with
$ -1/2 \leq \tau_1 \leq 1/2$~\cite{conform}. Therefore the case where 
$\tau_1$ is equal to an integer is equivalent to $\tau_1=0$, and the
case where $\tau_1$ is equal to a half integer is equivalent $\tau_1=1/2$.

Now, since $\tau_1=n/K$, we find that $\tau_1=1/2$ implies $K=n/\tau_1=2n$,
where n is an integer since $n_2+n_2'=n$. We conclude that if $\tau_1=1/2$ we 
cannot formulate a two-particle state for all integer values of $K$.
It is unacceptable to forbid some basis states at some resolutions; therefore, 
we conclude that we cannot formulate standard DLCQ on a torus with 
$\tau_1=1/2$.  Thus we find that it is only possible to form two-particle 
basis states on a torus with a shape angle if $\tau_1$ is an integer.  This
is equivalent by modular invariance to $\tau_1=0$, which is standard DLCQ 
without a shape angle. We conclude that for the allowed tori with a shape 
angle the physics is equivalent to standard DLCQ.

It is interesting to take an explicit look at modular invariance and show that 
the free energy of the two-particle state with $\tau_1=1$ is equivalent to the 
case $\tau_1=0$.   The free energy of a general two-particle state is obtained 
by applying the Hamiltonian to $|\psi_2\rangle$.  We find
\begin{eqnarray}
P^-|\psi_2 \rangle&=&\sum_{n+=1}^{K-1}\sum_{n_2=-T_l}^{T_l+2n_+\tau_1}
\frac{R_+}{2}\left(\frac{1}{n_+} +\frac{1}{K-n_+}\right)
\left( m^2 +\frac{1}{R_2^2\sin\theta^2} (n_2 - n_+\tau_1)^2 \right )
\nonumber \\
&&\times f_2(n_+,n_2) A^\dagger(n_+,n_2) A^\dagger(K-n_+,n-n_2) |0\rangle\,.
\end{eqnarray}
Now, if we take $\tau_1=1$ and therefore $n=K$, we find from
Eq.~(\ref{k2}) that
\begin{equation}
k_2=\frac{1}{R_2 \sin\theta}(n_2-n_+)\,, \quad
k_2'=-\frac{1}{R_2 \sin\theta}(n_2-n_+)\,.
\label{k22}
\end{equation}
We next make a change of variables to $p_2\equiv n_2-n_+$. It is appropriate 
to also relabel the creation operators with $p_2$ and $-p_2$ according to
Eq.~(\ref{k22}). We define a new integer, $T\equiv T_l+n_+$, to be used in 
the limits of the transverse sum. We then obtain
\begin{eqnarray}
P^-|\psi_2 \rangle&=&\sum_{n+=1}^{K-1}\sum_{p_2=-T}^{T}
\frac{R_+}{2}\left(\frac{1}{n_+} +\frac{1}{K-n_+}\right)
\left( m^2 +\frac{1}{R_2^2\sin\theta^2} p_2^2 \right )\nonumber \\
&&\times f_2(n_+,p_2) A^\dagger(n_+,p _2) A^\dagger(K-n_+,-p_2) |0\rangle\,.
\label{dlcq}
\end{eqnarray}
After rescaling $R_2$ by $\sin\theta$, we find as expected the standard 
DLCQ result for the free energy of a two-particle system. This argument can be
extended to systems with higher numbers of free particles.

\section{SDLCQ} \label{sec:SDLCQ}
Now let us consider the interacting theory ${\cal N}=1$, SYM theory
in 2+1 dimensions. This is a theory we that have solved on a rectangular 
torus using SDLCQ~\cite{hpt,Haney:2000tk,Antonuccio:1999zu}.
SDLCQ is a numerical method that exactly preserves the supersymmetry 
and therefore renders this theory totally finite. The only real
difference between DLCQ and SDLCQ is that the supercharge $Q^-$ is constructed 
in the Fock basis and the Hamiltonian is constructed by squaring the 
supercharge.  The Fock basis is the same as DLCQ, and the
arguments in the previous section follow essentially unchanged. 

The supercharge for ${\cal N}=1$ SYM theory has a rather simple form.  It is
\begin{equation}
Q^-=Q^-_{||}+\sum_{n_2} \frac{n_2}{R_2} Q^-_2\,.
\end{equation}
The transverse momentum explicitly appears in only one location. When we 
quantize the theory on the torus with a shape angle, we find
\begin{equation}
Q^-=Q^-_{||}+\sum_{n_2} \frac{(n_2-n_+\tau_1)}{R_2\sin\theta} Q^-_2\,.
\end{equation}
The operators $Q^-_{||}$ and $Q^-_2$ have the same form as on the rectangular 
lattice, except that they are written in term of the Fock operators of the
lattice with the shape angle.  

As in the free case, a symmetric cutoff on the transverse momentum in the 
TMCM frame requires that $\tau_1$ be an integer.  From modular invariance this 
is of course equivalent to the rectangular torus of standard SDLCQ. We can 
explicitly demonstrate this equivalence by shifting the transverse sum and 
rescaling $R_2$, exactly as we did in the free DLCQ case, to reproduce the
quantized supercharge found for the rectangular torus.

\section{Summary} \label{sec:Summary}

We considered standard DLCQ for a free scalar theory and standard SDLCQ for 
${\cal N}=1$ SYM theory, in 2+1 dimensions,  compactified on a rectangular torus 
and on a torus with a shape angle $\theta$. The ``standard'' definition 
uniquely defines these cutoff theories so that in the comparison we can 
be assured that we are only looking at the effect of the shape angle of the 
torus.  We find that it is only possible to formulate these theories with
transverse parity as a symmetry on tori with $\tau_1$ equal to an integer. 
Modular invariance then shows that tori with integer $\tau_1$ are equivalent 
to the rectangular torus, and therefore the physics of standard DLCQ and
SDLCQ are unchanged on the allowed tori with a shape angle.

A possible direction for future work would be to consider the implications 
of longitudinal parity\cite{Burkardt:1996gp} on this problem.

\section{Acknowledgments}
This work was supported in part by the U.S. Department of Energy. The author
would like to thank Keith Dienes and John Hiller for helpful discussion on 
this research.


\end{document}